\documentclass[aps,prd,superscriptaddress,12pt]{revtex4-1}
\usepackage{amsmath,amssymb,amsfonts,amsthm,dsfont,dcolumn}
\usepackage{graphicx,wrapfig}
\usepackage{tensor}
\usepackage{color}

\begin{document}

\begin{titlepage}
\title{A (gravitational) toy story}
\author{W. Barreto}
\affiliation{Departamento de F\'\i sica Te\'orica, 
Instituto de F\'\i sica A. D. Tavares,
Universidade do Estado do Rio de Janeiro, 
R. S\~ao Francisco Xavier, 524,
Rio de Janeiro 20550-013, RJ, Brasil}
\affiliation{Centro de F\'\i sica Fundamental, Universidad de Los Andes, 
M\'erida 5101, Venezuela}
\author{H. P. de Oliveira}
\affiliation{Departamento de F\'\i sica Te\'orica, 
Instituto de F\'\i sica A. D. Tavares,
Universidade do Estado do Rio de Janeiro, R. S\~ao Francisco Xavier, 
524, Rio de Janeiro 20550-013, RJ, Brasil}
\author{B. Rodriguez-Mueller}
\affiliation{Computational Science Research Center, 
San Diego State University, United States of America}

\begin{abstract}
\date{\today}
Frequently in Physics, insights and conclusions can be drawn from simple, idealized models. The discovery of critical behavior in the gravitational collapse of a massless scalar field leads to the simulation of binary black holes, from its coalescence to merging and ringdown. We refined a toy model to explore black hole formation as these events unfold to revisit the instability of a gravitational kink. We confirmed a conjecture related to a mass gap, for critical behavior at the threshold of black hole formation. We find a critical exponent twice the standard value. Surprisingly, this larger critical exponent is also present in the multiple critical behavior for the black hole formation from a massless scalar field in asymptotically anti-de Sitter spacetimes. What is the meaning of this mass gap? Does it have physical relevance? 
 
\end{abstract}
\maketitle

\end{titlepage}

\begin{flushright}
{``{\it After more than a hundred years, \\
the details are irretrievable, \\
but it is not hard to conjecture what happened.''
\\ Jorge Luis Borges,
\\ The garden of forking paths,\\ Fictions (1944).} 
}
\end{flushright}

The most beautiful theory is now a hundred years old. Around the time of its 100th birthday, echoes of a remarkable event provided a signal that confirmed it yet again. Evasive, incredibly weak, and almost undetectable gravitational waves were finally detected. These results give us another window into the universe and allow us to continue understanding nature. Many open questions remain in gravity, and the key one is: How to merge the gravitation and quantum theories? Critical behavior is a good place to find edge cases that may shed light on the relation of general relativity to quantum field theory, as in the context of AdS/CFT dualities. 

The scalar field is an excellent toy model. If the underlying physics is fundamental, the mathematical structure of the field equations and theoretical discoveries may lead to realistic situations. A massless scalar field allows the evolution of non-topological solitons, kinks, black holes from gravitational collapse and can show the onset of chaos and turbulence. Here we tell a story about a topological kink.

A self-gravitating massless scalar field has been crucial for general relativity in different ways. Black hole coalescence and merging predicts expected profiles of gravitational radiation, relevant for detection \cite{pretorius}, \cite{campanellietal}, \cite{bakeretal}, \cite{abbotetal16}. A connection with dark matter has been speculated (see Ref. \cite{darkm} and references therein), also in this context, kinks are used to model topological defects \cite{bhmdodds10}. The self-gravitating spherical and massless scalar field described by Einstein-Klein-Gordon (EKG) has the same mathematical structure as the non-spherical field equations in vacuum \cite{winicour_liv}. Thus, we can reasonably say that the scalar field mimics gravitational radiation. 

While working on the strong field limit near the formation of a black hole, Choptuik found Type II critical behavior\cite{c93} for spherically symmetric massless scalar fields minimally coupled to gravity. Type I critical behavior is observed when a massive scalar field is considered \cite{brady}, \cite{ss}. A review of critical phenomena for gravitational collapse is available in Ref. \cite{gm}, including quantum versions. 

Despite extensive study of gravitational critical behavior with massless scalar fields, there is plenty of exploration left about the non-linear nature of gravitation. For Type II critical behavior, is the final mass infinitesimal or finite? \cite{psa05}. For the black hole case, the existence of a mass gap associated with a power-law scaling is not clear \cite{mc}, \cite{cd15}. What happens between a kink collapse and the formation of a black hole? \cite{bglw96}. Before diving into these questions, it might be interesting to look back at the idea of the kink.

Wheeler formulated the original idea of solitons in general relativity with the concept of geons \cite{wheeler}. Wheeler's geon is a spherical shell of electromagnetic radiation held together by its gravitational attraction. To a certain extent, Wheeler's geon anticipated the topological and non-topological solitonic solutions of classical nonlinear field theories \cite{rajamaran}, whose main 
initial motivation was to model elementary particles. In this instance, we remark that the nonlinearity of the field equations like those of general relativity is the crucial ingredient to produce solitonic structures. 

Further theoretical development has uncovered a large variety of soliton solutions composed of distinct matter fields including the idealized thin spherical geon of Pfister and Braun \cite{pfister}, solitons with toroidal electromagnetic waves \cite{ernst}, cylindrical geons \cite{melvin}, neutrino geons \cite{brill_wheeler} and pure gravitational solitons made of gravitational waves \cite{brill_hartle}.  These structures, being stable or unstable, can contribute to dark matter. 

In particular, we mention here two spherically symmetric models of gravitational solitons or kinks made of self-gravitating scalar fields. The first model is due to Kodama \cite{kodama1}, a self-gravitating repulsive scalar field with a $\lambda \phi^4$ potential. The resulting configuration is static, singularity-free and has finite energy. Kodama has named it as the general relativistic kink since it might be considered a generalization of the usual one-dimensional kink solution. Later, Kodama {\it et al.} \cite{kodama2} have proven the stability of the relativistic kink under small radial perturbations. An interesting question, not investigated yet (and not addressed here) is under which conditions the collapse of a scalar field could end up in the general relativistic kink. 

The second model is due to Barreto et al. \cite{bglw96} and consists of a static and asymptotically flat massless scalar field that has a fixed value or potential at an interior mirror located at $r=R$, with $r$ the usual radial coordinate. In this case, $\varphi(R)=A$, and the spacetime in the interior of the reflector is considered flat. Therefore, the configuration has typical kink boundary conditions. Also, it represents an extremum of the energy, subject to a fixed potential at the interior mirror. Barreto {\it et al.} have shown that the static configuration can be stable or unstable exhibiting a turning point instability about the critical $A_c$ value of the potential. 

Next, we discuss the role of spherically symmetric kink structures as the result of a gravitational collapse of scalar fields, based on our recent work \cite{bcdrr16}. More important is the connection with critical phenomena in the gravitational collapse. 

We consider the following line element in Bondi coordinates \cite{EKG}, \cite{X1986}, 

\begin{eqnarray}
d s^2 = -\frac{V}{r}\mathrm{e}^{2\beta} du^2 - 2\mathrm{e}^{2\beta}du dr + r^2 (d\theta^2+\sin^2 \theta d\phi^2),\nonumber \\
 \label{eq1}
\end{eqnarray}

\noindent where $u$ is the asymptotically retarded time, $r$ is the radial coordinate and the metric functions $V$ and $\beta$ depend on $u,r$. The relevant field equation reduce to the hypersurface equations, 

\begin{eqnarray}
\beta_{,r} &=& 2 \pi r \varphi_{,r}^2 \label{eq2},\\
\nonumber\\
V_{,r} &=& \mathrm{e}^{2 \beta} \label{eq3}
\end{eqnarray}

\noindent and to the scalar wave equation,

\begin{equation}
2(r \varphi)_{,ur} - \frac{1}{r}(r V \varphi_{,r})_{,r} = 0. \label{eq4}
\end{equation}
\noindent To evolve the field equations we have established the initial data $\varphi_0(r)=\varphi(u_0,r)$ for $r \geq R$ that is the radius of the mirror. At the mirror, we set $\varphi(u,R)=A=\mathrm{constant}$, and also, the gauge condition $\varphi(u,\infty)=0$. To guarantee a unique evolution, additional conditions must be added: the coordinate condition $\beta(u,R)=0$, and $V(u,R)=R$ for a continuous match to a flat interior spacetime for $0 \leq r \leq R$.

A static solution \cite{jnw1968}, by Janis-Newman-Winicour (JNW), exists in null coordinates when $\varphi_{,u}=0$ in the wave equation (\ref{eq4}), giving us
\begin{equation}
	  rV \Psi_{,r} = \text{constant},  \label{eq:lstat}
\end{equation}
then using (\ref{eq2}) and (\ref{eq3}) to
solve the $r$-dependence, we arrive to the solution: 
\begin{equation}
\Psi(V) = {1 \over {4\sqrt{\pi} \>{\cosh} \alpha}} \ln
        \left[{{V + R\> (e^{2\>\alpha} - 1)} \over
         {V + R\> (e^{-2\>\alpha} - 1)}} \right] ,\label{eq:static}
\end{equation}
where $r(V)=r_\Psi$ is given by
\begin{eqnarray}
        r_\Psi^2 &=&  e^{-4\> \alpha \tanh \alpha}
               [V + R\> (e^{-2\>\alpha} - 1)]^{1\> -\> \tanh \alpha}\times
               [V + R\> (e^{2\>\alpha} - 1) ]^{1\> +\> \tanh \alpha}.
                \label{eq:rstatic}
\end{eqnarray}
The spacetime has a naked singularity when it is extended analytically to
$r=0$~\cite{jnw1968}.

The asymptotically flat static solution of the EKG system is an extremum of the energy, subject to a fixed kink potential at $r=R$ \cite{bglw96}. Here the integration constant $\alpha$ determines the kink potential {(its amplitude at $r=R$)}, 
\begin{equation}
       A_\Psi(\alpha)={\alpha \over \sqrt{\pi}\cosh\alpha}. \label{eq:kink}
\end{equation}
{
For this solution, the Bondi mass is
}
\begin{equation}
    M_{\Psi}(\alpha)=2R\sinh^2\alpha e^{-2\alpha\tanh\alpha}.
             \label{eq:statm}
\end{equation}

We build three types of initial kinks:
\begin{itemize}
\item{\it Kink I.} For numerical testing and convergence we use the noncompact initial kink 
\begin{equation}
\varphi(0,r)=\frac{2(A_c+\lambda)R}{R+r},
\end{equation}
where $A_c$ is the critical amplitude for the static kink ($\approx 0.3656$), corresponding to the turning point \cite{bglw96}; $\lambda$ is a parameter to control the amplitude. For a kink potential $A > A_c$ no static equilibria exist. Any initial state undergoes a black hole formation. For a critical value of $\lambda^*=0$ no mass gap exists and we have to wait an infinite proper time to observe whether or not a black hole is formed.  
\item {\it Kink II.} For this kink, we prescribe a value 
of $\alpha$, and then truncate the static solution (\ref{eq:static}) at $r=R$, then adding a global scale perturbation given by
\begin{equation}
\varphi(0,r;\alpha)=\Psi + \lambda\left[ \frac{R-r}{2r(r+R)} \right],
\end{equation}
with static kink values as boundary conditions. When $\alpha=1$, a critical point is found at $\lambda^*\approx0.1929$. 
Below $\lambda^*$, the perturbed kink does not result in a black hole; instead, decays to the static solution \cite{bglw96}. 
Above $\lambda^*$ the system always forms a black hole. 
\item {\it Kink III.} From numerical experimentation we arrived to the following kink
\begin{equation}
\varphi(0,r)=\frac{R(A_c+\lambda)}{re^{(r-R)^2/\sigma^2}},
\end{equation}
with a variance $\sigma=1/2$. The Gaussian-like shape makes this kink useful to explore the critical behavior in the gravitational collapse. 

\end{itemize}

\begin{figure}[!ht]
\begin{center}
\includegraphics[width=4.in,height=2.5in]{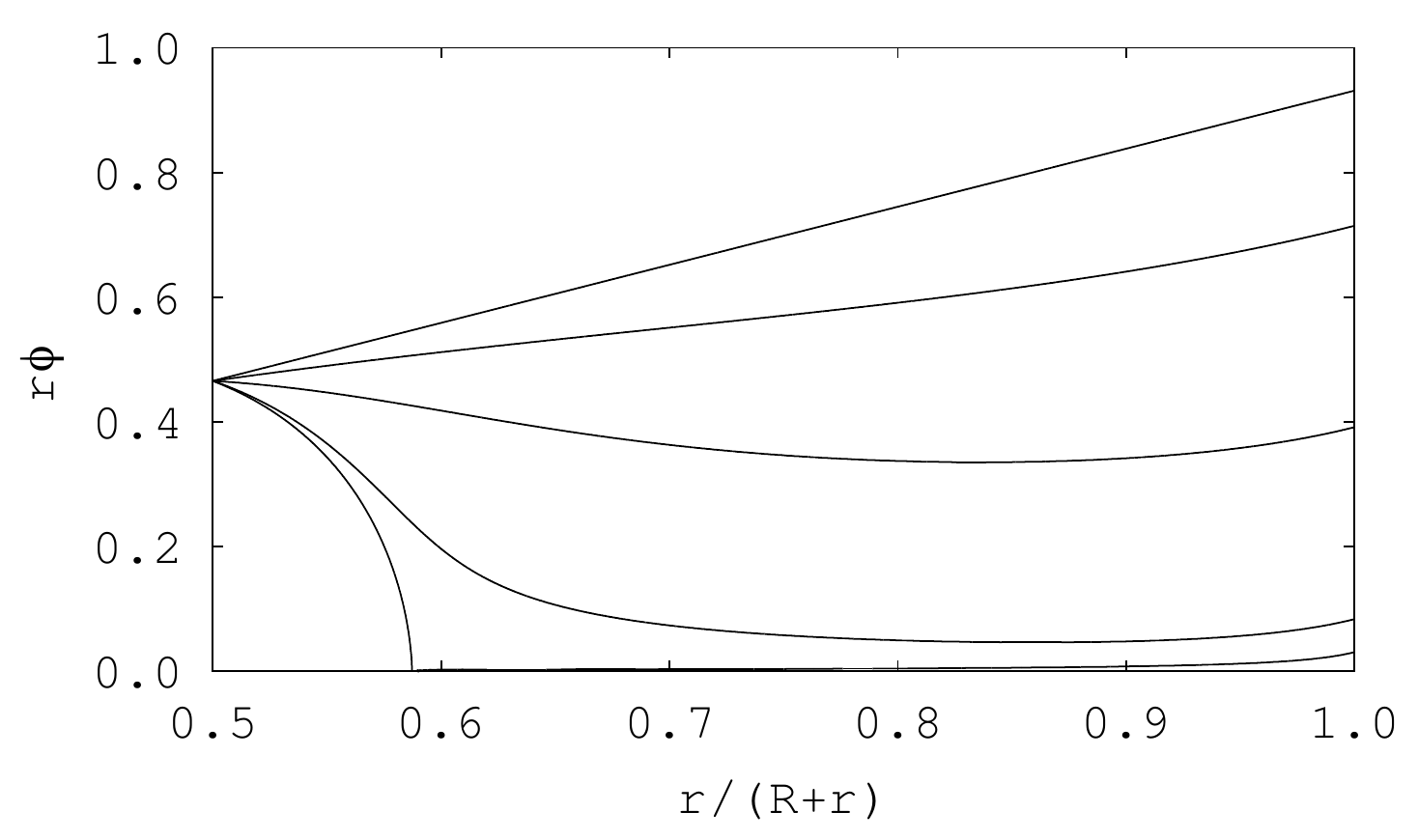}
\caption{Evolution of the massless scalar field $r\varphi$ as a function of the compactified radial coordinate $r/(R+r)$ for Kink I, with $\lambda=0.1$ for a grid size of $\approx 8,000$. The upper curve represents $u=0$; and the lower curve, $u\approx 3.45$. The system collapses to a black hole shedding the ``hair'' of the exterior field, as described in \cite{bglw96}. For the whole evolution, we see the ``conservation'' of the Newman-Penrose constant. The slope $r^2\partial_r (r\varphi)$ at $r/(R+r)\approx 1$ is conserved up to the black hole formation within $0.003\%$ of relative percentual variation on the initial value of the Newman-Penrose constant. This test provides evidence of numerical convergence.}
\label{evolution}
\end{center}
\end{figure}

\begin{figure}[!ht]
\begin{center}
\includegraphics[width=4.in,height=2.5in]{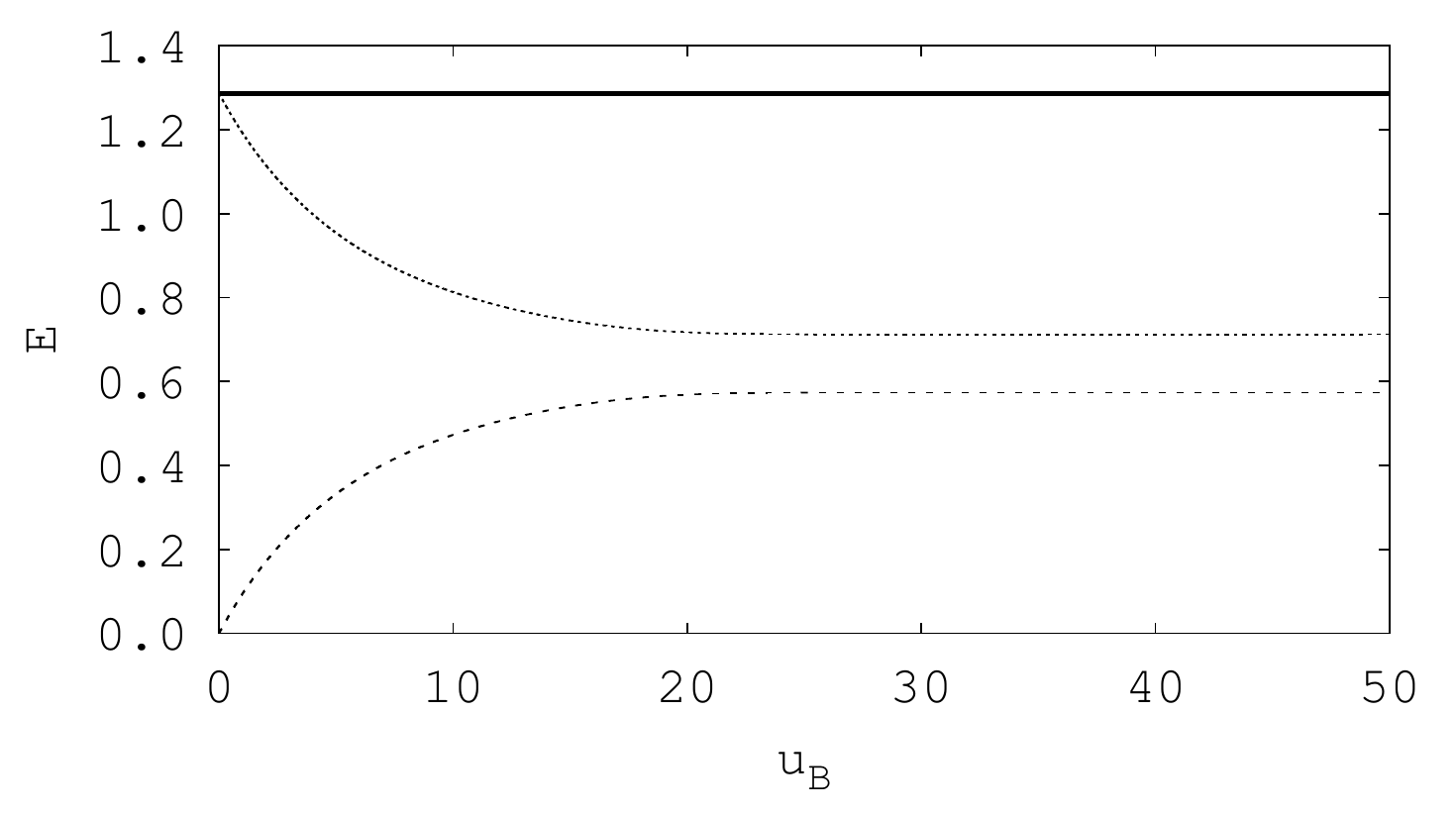}
\caption{Energy conservation as a function of the Bondi time, up to the black hole formation, as shown in Fig. 1. The descending curve corresponds to the Bondi mass; the ascending line corresponds to the energy flow to infinity; the horizontal curve corresponds to the algebraic sum of both curves. The system collapses to a black hole; for the whole evolution, we see the conservation of energy within $0.02\%$ of relative percentual variation on the initial value of the energy. This test provides additional evidence of numerical convergence.}
\label{evolution}
\end{center}
\end{figure}

\begin{figure}[!ht]
\begin{center}
\includegraphics[width=4.in,height=2.5in]{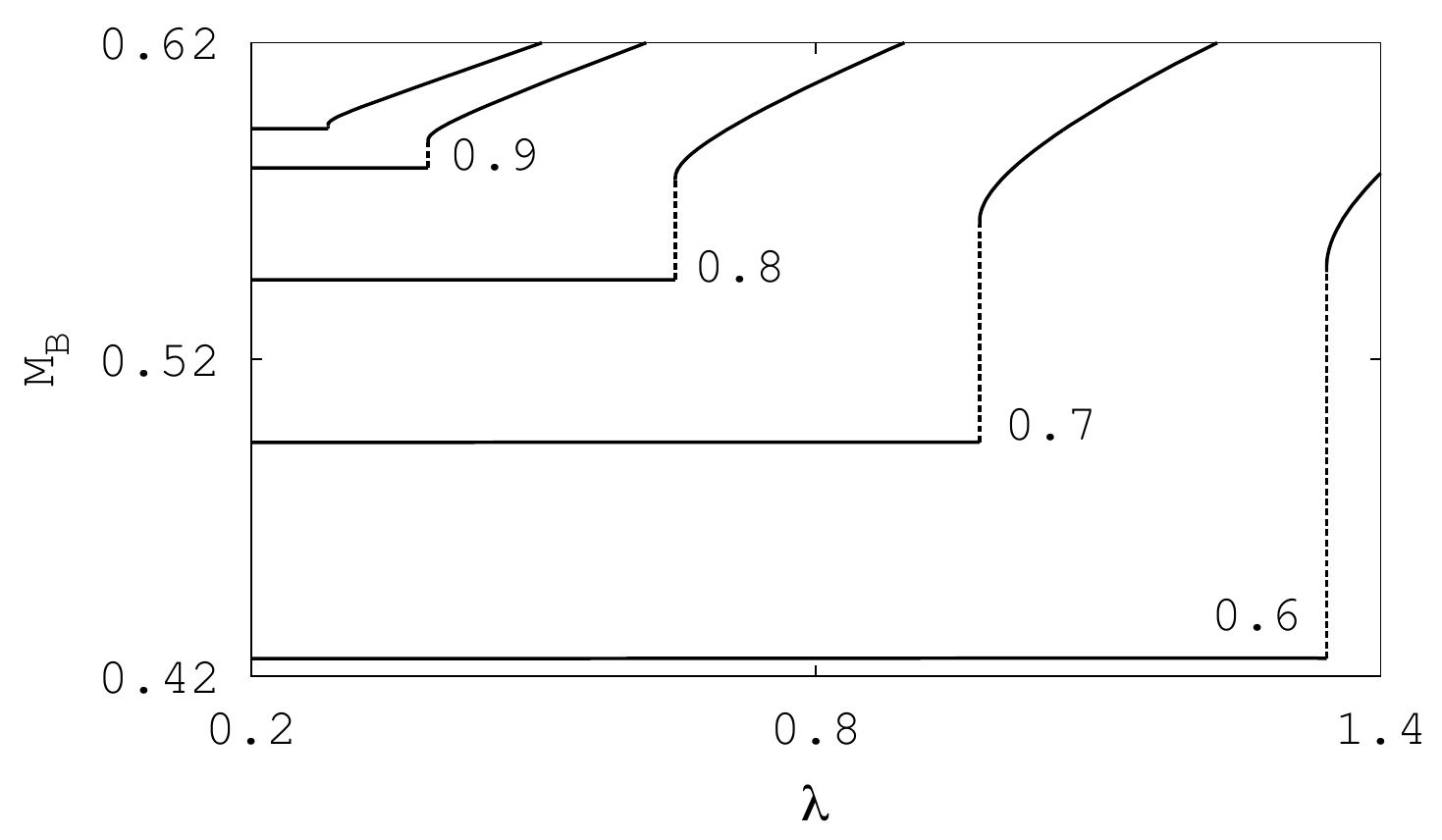}
\caption{The Bondi mass $M_{B}$ as a function of the parameter
$\lambda$ for different values of $\alpha$ in Kink II: 
$0.6$; $0.7$; $0.8$; $0.9$ and $0.95$ (curve not labeled in graph).
For each  $\alpha$ we have a critical value $\lambda^*$ and a mass gap.
The mass gap closes or becomes really small near $\alpha\approx 0.95$.}
\label{massgapII}
\end{center}
\end{figure}

\begin{figure}[!ht]
\begin{center}
\includegraphics[width=4in,height=2.5in]{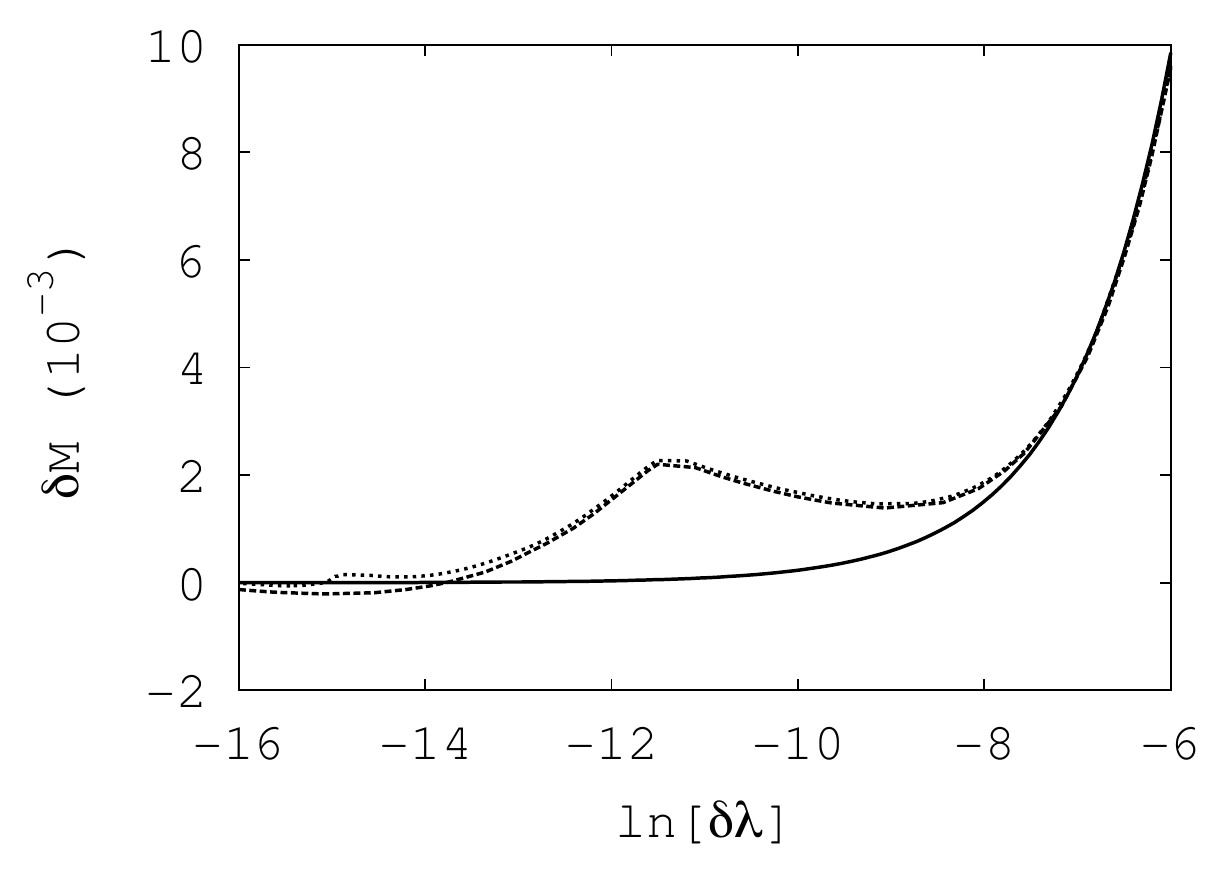}
\caption{$\delta M$ as a function of $\ln(\delta\lambda)$ in Kink III (dotted line). The continuous curve roughly grows up with a power of $2\gamma\approx0.74$. From the oscillatory main component we get a period of $\Delta/2\gamma\approx 4.59$. Other unstable modes of decaying oscillations are apparent. In this graph, we also show results using the Galerkin-Collocation method (dashed line). Both methods, finite differences, and Galerkin-Collocation give the same results up to some resolution, giving us confidence in the numerical results.}
\label{critical}
\end{center}
\end{figure}

\begin{table}[!ht]
\begin{ruledtabular}
\begin{tabular}{|l|c|c|c|c|c|c|c|c|}
 &coarse&medium&fine&&coarse&medium&fine& \\
\hline
$u_B$ & $M^c_B$  & $M^m_B$  & $M^f_B$ & $n_{M_B}$ & $C^c_{NP}$ & $C^m_{NP}$ & $C^f_{NP}$ & $n_{C_{NP}}$\\
\hline
1.0 &    1.19209 &  1.19190 &  1.19180   &    0.99  & 0.9312502 & 0.9312511 &  0.9312513 & 2.00 \\
\hline
2.0  &   1.11554 &  1.11520 &  1.11502   &    1.00  & 0.9312488 & 0.9312507 & 0.9312511 & 2.00 \\
\hline
3.0  &  1.05203  &  1.05174 & 1.05160  & 1.00 & 0.9312469 & 0.9312503 & 0.9312511 &  2.00    \\
\end{tabular}
\end{ruledtabular}
\caption{Cauchy rate of convergence, $n=\log_2\{(X^c-X^m)/(X^m-X^f)\}$, where $X$ represents the Bondi mass ($M_B$) or the Newman-Penrose constant ($C_{NP}$), at different Bondi times, $u_B$, for a: coarse grid (c; grid size of 2,000); medium grid (m; grid size of 4,000); fine grid (f; grid size of 8,000). The initial datum is the Kink I, as in figures 1 and 2.}
\label{tab:derivatives}
\end{table}

For calculations in this context, we use two well established numerical schemes. We use the finite differences and Galerkin-Collocation methods for the characteristic formulation \cite{gwi}, \cite{gw}, \cite{drs}. Also, we use scripting to explore the critical point near the bifurcation, running and analyzing a suitably large number of models \cite{bcdrr16}.

Figure 1 shows the evolution of the Kink I up to the black hole formation, with $\lambda=0.1$ for a grid size of $\approx 8,000$. The ``conservation'' of the Newman-Penrose constant (slope of the kink at null infinity) is apparent. The relative percentual variation of the initial value of the Newman-Penrose constant is about $0.003\%$. 

Figure 2 displays the energy conservation for the same evolution as in Fig. 1 (up to the black hole formation). The relative percentual variation on the initial value of the conserved energy is about $0.02\%$. 

Table I shows an evidence of convergence, in the sense of Cauchy, to first order for the Bondi mass and second order for the Newman-Penrose constant. We use as criterium for the black hole formation a gravitational redshift of the order of $10^7$ which coincides with the apparent horizon formation. 

Figure 3 displays the expected mass gap as a function of $\lambda$ for different values of $\alpha$ in Kink II. To obtain these results we select a static solution ($\alpha$) and vary the parameter $\lambda$ of the global perturbation, we evolve and find each critical value for which the black hole forms or not. We then show a spectrum with mass gap branches. For the black hole mass, we approximated with the minimum value reached by the Bondi mass.


Figure 4 shows the mass spectrum in the supercritical case for Kink III, that is when a black hole always forms. For this particular graph, we have considered the Galerkin-Collocation method, which gives us numerical confidence on the obtained results. To get this figure, we first evolve the system up to the black hole formation and estimate the critical pair $(M^*,\lambda^*)$. Then we evolve again up to the black hole formation only for $\lambda > \lambda^*$ to obtain a spectrum $(M_{BH},\lambda)$. Thus, from numerical experimentation we infer the following mass scaling power law
\begin{equation}
\delta M=K\delta\lambda^{2\gamma} + f [K\delta\lambda^{2\gamma}],
\end{equation}
where $\delta M=M_{B}-M^*$, $\delta\lambda=\lambda-\lambda^*$, $f$ being a non-trivial function of its argument, and $K$ a fitted constant. $M^*$ is the supercritical mass limit which corresponds to the critical amplitude $\lambda^*$. Each point for the black hole spectrum $(M_{BH}; \lambda)$ consumes about 65 minutes (grid size $\approx 15,000$) using a N1-standard-1 virtual machine on the Google Compute Engine. 

{The gravitational collapse of the scalar field produces a horizon. Some energy radiates away to infinity; the rest goes to the interior contributing to the final black hole mass. It is clear that the mirror falls in through the horizon. To some extent, this problem is the same critical behavior investigated by Choptuik \cite{c93}. But now we have a mass gap because the final mass of the black hole is greater than $R/2$, containing the mirror. Figures 3 displays the expected mass gap. We would like to reiterate that the black hole mass can be finite in the critical collapse of Type II. The new features of this critical behavior are due to the kink setting, and there might be additional features awaiting discovery. Figure 4 displays interesting behavior, such as echoing and power law mass scaling.}

We emphasize here the finding of the expected mass gap conjectured in Ref. \cite{bglw96}. We know that the asymptotic symmetric group, of Bondi-Metzner-Sachs, is conjectured to be dual to a conformal field theory (see for instance \cite{bt10}). Then on physical grounds, we can expect that the mass gap introduced by the kink could correspond to some mass gap in other dimensionality and physical context, e.g., to the two-magnons-bound-state mass gap for the $(2+1)$ dimensional Ising model \cite{n14}. Finally, it was reported recently by Santos and Sopuerta a multiple critical behavior \cite{ss16}, \cite{ss16b}, in the context of the black hole formation from a massless scalar field in asymptotically anti-de Sitter spacetimes. Close to each critical parameter, there are two types of power scaling mass law: with a mass gap and without a mass gap. It is interesting the twice-fold critical exponent when the mass gap is apparent, as in the kink problem \cite{bglw96}, \cite{bcdrr16}.   

The authors thank the financial support of Brazilian agencies CNPq and FAPERJ.

\end{document}